\documentclass[
]{ceurart}

\usepackage{listings}
\usepackage{comment}
\usepackage{soul}
\usepackage{url}
\usepackage[utf8]{inputenc}
\usepackage[small,justification=centering]{caption}
\usepackage[inline]{enumitem}
\usepackage{array}
\usepackage{graphicx}
\usepackage[justification=centering]{subcaption} 
\usepackage[table,xcdraw]{xcolor}
\usepackage{amsmath}
\usepackage{amsthm}
\usepackage{booktabs}
\usepackage{algorithm}
\usepackage{algorithmic}
\usepackage[switch]{lineno}
\usepackage{makecell} 
\usepackage{multirow}
\usepackage{adjustbox} 

\begin{document}

\copyrightyear{2025}
\copyrightclause{Copyright for this paper by its authors.
  Use permitted under Creative Commons License Attribution 4.0
  International (CC BY 4.0).}

\conference{TRUST-AI: The European Workshop on Trustworthy AI. Organized as part of the European Conference of Artificial Intelligence - ECAI 2025. October 2025, Bologna, Italy.}

\title{Trust in Vision-Language Models: Insights from a Participatory User Workshop}

\author[A]{Agnese Chiatti}[%
orcid=,
email=agnese.chiatti@polimi.it
]
\cormark[1]

\author[C]{Lara Piccolo}[%
orcid=,
email=lara.piccolo@code.berlin
] 

\author[B]{Sara Bernardini}[%
orcid=,
email=sara.bernardini@cs.ox.ac.uk
]

\author[A]{Matteo Matteucci}[%
orcid=,
email=matteo.matteucci@polimi.it
]

\author[A]{Viola Schiaffonati}[%
orcid=,
email=viola.schiaffonati@polimi.it
]

\address[A]{Politecnico di Milano, Italy}
\address[B]{University of Oxford, UK}
\address[C]{CODE University of Applied Sciences, Germany}

\cortext[1]{Corresponding author.}

\begin{abstract}
With the growing deployment of Vision-Language Models (VLMs), pre-trained on large image-text and video-text datasets, it is critical to equip users with the tools to discern when to trust these systems. However, examining how user trust in VLMs builds and evolves remains an open problem. This problem is exacerbated by the increasing reliance on AI models as judges for experimental validation, to bypass the cost and implications of running participatory design studies directly with users. Following a user-centred approach, this paper presents preliminary results from a workshop with prospective VLM users. Insights from this pilot workshop inform future studies aimed at contextualising trust metrics and strategies for participants' engagement to fit the case of user-VLM interaction.
\end{abstract}

\maketitle

\section{Introduction}
Vision Language Models (VLMs) represent a methodological shift for learning correspondences between image-text and video-text pairs from large-scale data. Unlike traditional Computer Vision approaches, VLMs reduce reliance on curated, task-specific training datasets, enabling zero-shot inference on previously unseen tasks and categories \cite{zhang2024vision}. 
Thanks to their remarkable ability to interpret image and video content, these models are being rapidly adopted by society at large. While this unprecedented adoption presents exciting opportunities, it also raises significant concerns. VLMs are inherently difficult to audit, often closed-source, and operate as black-box systems accessible only through indirect observation, steering Artificial Intelligence (AI)  research closer to an ``ersatz natural science” \cite{kambhampati2024can}. 

These challenges are heightened when VLMs are used in safety-critical environments robotic systems, where errors in decision-making can lead to catastrophic consequences \cite{cummings2021rethinking} - 
like in the case of autonomous transportation, \cite{perez2024artificial},
inspection and maintenance \cite{jovan2023adaptive},
disaster response \cite{schiaffonati2016stretching},
and assistive healthcare \cite{chiatti2023visual}.
Human oversight in evaluating VLM performance in real-world settings is crucial to ensure users can determine how and when to trust these systems. 

The EU Ethics Guidelines for Trustworthy AI \cite{EUTAI} and AI Act \cite{EUAct} provide a regulatory framework grounded on ethical principles, varying levels of risk, and technical requirements. However, defining Trustworthy AI (TAI) within specific real-world contexts remains contested \cite{zanotti2023keep}. The first gap is epistemological, requiring a distinction between \textit{trustworthiness}, an inherent property of a system's actual capabilities, and \textit{trust}, which reflects the user's perception of trustworthiness \cite{mayer1995abi}. Moreover, the term \textit{trust}, borrowed from interpersonal contexts, has been directly applied to AI, yet further research and regulation are needed to adapt this notion to human-AI interactions. 

Adding to the complexity, the concept of trust overlaps with other key dimensions of Human-AI interaction, such as \textit{explainability} \cite{chander2024xai}, \textit{transparency} \cite{barman2024beyond}, \textit{fairness} \cite{calegari2023assessing}, and \textit{accountability} \cite{novelli2024accountability}. Thus, the lack of consensus around user trust in AI stems, at least in part, from the insufficient clarity regarding the socio-technical factors that shape trust and their complex interdependencies. 
\vspace{-1em}
\paragraph{Problem scope.} Trust dynamics are hard to define, as they are inherently subjective and dependent on the application context. For example, the level of trust placed in a VLM may differ when describing a video for entertainment versus one used in a medical diagnosis. Perceived trust also evolves over time, shaped by prior experience, familiarity with the technology, and repeated interactions \cite{mehrotra2024integrity}. Given the broad and multi-faceted nature of the field of TAI, our focus is on \textit{examining trust dynamics during users' interaction with Vision Language Models}.  As highlighted in a recent coverage study of works in user-VLM trust \cite{chiatti2025mapping}, the current literature lacks comprehensive studies directly involving users, especially when considering applications in Computer Vision and multi-modal AI models. In this paper, we draw on \textbf{insights from a workshop with users} to ground the notion of VLM trust in a concrete use case and gather preliminary requirements for future studies.

\section{Related work} 
\label{sec:lit_review}

\subsection{Vision Language Models (VLMs)}  
The Vision Language Model learning paradigm typically involves three stages. A \textit{pre-training} phase, where the model is optimised using large-scale, off-the-shelf data, either labelled or unlabelled. An optional \textit{fine-tuning} phase, where the model is adapted to domain-specific data. An \textit{inference} phase, where the model performance is evaluated on downstream tasks. VLMs have gained particular attention for their ability to skip the fine-tuning step, enabling zero-shot inference on unseen tasks or categories without additional training \cite{zhang2024vision}. Zero-shot capabilities in VLMs arise from their design, as visual and textual features are extracted via encoder modules in a general-purpose manner, i.e., independently of any specific downstream task. Visual features are typically extracted from images and video frames using either Convolutional Neural Networks (CNNs) or Transformers, while textual features are almost invariably extracted with Transformers. 

Correlations between visual and textual features are learned through various pre-training objectives. These include i) \textit{contrastive objectives}, which optimise embeddings for positioning similar features closer and dissimilar features farther in the vector space; 
ii) \textit{generative objectives}, where correlations are learned by generating data within a single modality (e.g., image-to-image generation \cite{luo2023segclip}) 
or across modalities - e.g., text-to-image generation \cite{singh2022flava}; iii) \textit{alignment objectives}, which focus on directly matching corresponding elements in visual and textual inputs (e.g., matching local image regions to words \cite{li2022glip}). 
In the case of videos, models are often trained on iv) \textit{temporal objectives}, like re-ordering input frames \cite{chen2023vlp}.   

Earlier VLMs used separate branches for each modality during pre-training as in the original CLIP model \cite{radford2021learning}. However, recent architectures have shifted to unified designs with a single encoder for both modalities \cite{jang2023singletower,singh2022flava} to enable cross-modal feature fusion and reduce computational overhead \cite{zhang2024vision}.  

\subsection{User Trust in VLMs} 
Recent surveys have examined multiple factors influencing user trust in AI, including: (i) explainability \cite{chander2024xai}, (ii) model and data fairness \cite{calegari2023assessing}, (iii) robustness \cite{chander2024xai,tocchetti2024robustness}, and (iv) accuracy, often framed in terms of trade-offs among these dimensions \cite{li2024triangular}. Additional studies investigate trust across the AI lifecycle \cite{calegari2023assessing}, in software development \cite{liu2024tertiary}, and in domains such as autonomous robotics and safety-critical systems \cite{methnani2024who,perez2024artificial}. While these works provide valuable overviews agnostic to specific models or modalities, fewer studies have focused on trust and trustworthiness in Computer Vision.

Recently, several conceptual frameworks have emerged to formalise trust in Large Language Models (LLMs), such as the comprehensive TrustLLM framework \cite{huang2024trustllm}. However, extending the investigation to VLMs requires also to consider the components of visual perception and reasoning that contribute to trust building. 
To examine trust in the context of mixed human-VLM teams pursuing shared goals, in Chiatti et al. \cite{chiatti2025mapping}, we extended the seminal ABI model of trust in organisations \cite{mayer1995abi}, integrating insights from Situated Cognition \cite{collins2024building,lake2017building} and Theory of Mind \cite{verma2024tom}. Building on the three ABI dimensions, we characterised VLM Ability using core components of intelligent systems as defined by Lake et al. \cite{lake2017building}, further linked to cognitive science motifs in Collins et al. \cite{collins2024building}. Benevolence was categorised through collaborative thought modes described in Collins et al. \cite{collins2024building}, contextualised for human-VLM cooperation. For Integrity, agent behaviour types from Verma et al. \cite{verma2024tom} were employed to represent adherence to cooperative principles. The resulting taxonomy is available in the full survey paper \cite{chiatti2025mapping}. Here, we summarise key observations from mapping the existing literature on user-VLM trust to this taxonomy.
\vspace{-8pt}
\paragraph{Domain complementarity.} A notable gap in current research lies in the division of focus: TAI studies involving users predominantly address language-based modalities, while Computer Vision research prioritizes task-specific performance metrics over trust-related properties. Research on VLM trust - both in terms of methods and datasets - has primarily concentrated on mitigating hallucinations \cite{yu2024rlaif,sahu2024pelican,chen2024dress,huang_opera_2024,leng_mitigatingVCD_2024,deng_seeing_2024,fang_uncertainty_2024,fan_contextcam_2024,xiao2024can} and defending against adversarial attacks \cite{vatsa2024adventures,liu2024safety,gou2024eyes,khan_consistency_2024,islam_malicious_2024,xu_shadowcast_2024}. However, such efforts offer limited benchmarks for broader cognitive capabilities relevant to trust formation. 
Other integrity-related issues well-documented for LLMs \cite{huang2024trustllm} - including preference bias, data leakage, accountability flaws, emotional awareness, ethical dilemmas, decision opacity, and rhetorical manipulation - remain largely unexplored in VLM research. 
\vspace{-8pt}
\paragraph{Graphs as a bridge between language and perception.} In video understanding and situated reasoning tasks, benchmarking intent and action detection capabilities emerges as a core trend. Recent benchmarks - including ReX-Time \cite{chen2024rextime}, VHELM \cite{lee_vhelm_2024}, STAR \cite{wu2021star}, and AGQA \cite{grunde2021agqa} - highlight scene graphs as a promising representation for connecting people, objects, and events. In scene graphs, nodes denote objects (e.g., ``person'', ``table'') or attributes (e.g., ``red'', ``wooden''), while edges represent relations (e.g., ``on top of'', ``holding'', ``next to''). 
Crucially, this representation format can support semantic concept abstraction for meta-learning, i.e., learning to learn, \cite{wu2021star},  provide ``means-to-an-end'' representations of video narratives \cite{chen2024rextime}, and help structure model outputs and reasoning chains \cite{mitra2024compositional}, analogous to Chain-of-Thought prompting \cite{besta2024graph,lusha2025graph}. Notably, graphs offer a hybrid format: decomposable into text while grounded in compositional visual elements. Exploring this dual representation is particularly relevant given VLMs known over-reliance on language over perceptual modalities \cite{gou2024eyes,huai_debiased_2024}.
\vspace{-8pt}
\paragraph{Human-in-the-loop or Model-in-the-loop?} Studies on VLM trust and trustworthiness rarely involve users directly. This limitation is particularly severe considering the centrality of human oversight towards developing trustworthy AI systems, as also prescribed in the EU Ethics guidelines for TAI. Some adopt Human-in-the-Loop approaches, refining predictions through collaborative deliberation or aligning models with human feedback through collaborative learning, as seen in PIVOT \cite{nasiriany2024pivot} and RLAIF-V \cite{yu2024rlaif}. Nonetheless, the field remains dominated by "Model-in-the-Loop" methods \cite{khan_consistency_2024}, presumably due to the logistical complexity, ethical implications, and cost of user studies. Notable exceptions include Fan et al. \cite{fan_contextcam_2024}, who examine user-VLM collaboration in image co-creation, and Mehrotra et al. \cite{mehrotra2024integrity}, who assess user willingness to delegate calorie estimation from food images to VLMs via gamification. 

Given the scarcity of user-centred analyses of trust building in interactions with VLMs, this paper presents an exploratory workshop involving experts in Design and Development to elicit preliminary requirements for future studies. Drawing from our review of related work, the workshop focused on complex evaluation tasks requiring situational reasoning capabilities \cite{wu2021star}, and probed the potential of graph-based representations as interactive interfaces between users and VLMs. Following previous work by Mehrotra et al. \cite{mehrotra2024integrity}, in the workshop we modelled trust as task delegation (Section 3.1).

\section{User trust in VLMs: a case study} 
Evaluating trust dynamics requires a user-centred perspective. However, user-centred studies are inherently complex, as they require aligning the rapid advancements of technical solutions with participants' subjective understanding of state-of-the-art capabilities and limitations to ensure scientifically rigorous results. A pilot case study on user-VLM trust, albeit preliminary, can offer valuable insights and lessons learned to the research community and serve as a foundation for future studies involving a larger participant base, helping mitigate their complexity, risks, and costs. Furthermore, these preliminary findings can inform the design of a digital tool to support evaluations of user interactions with VLMs. 

To this aim, we organised an exploratory workshop guided by two main research questions. First, we sought to explore \textit{how trust dynamics develop when users engage with a VLM to solve collaborative planning and sensemaking tasks that require situated cognition abilities} (\textbf{RQ1}). Second, we aimed to gather expert feedback on \textit{which features should be prioritized in the design of a Web App to evaluate these trust dynamics in user studies} (\textbf{RQ2}).
\vspace{-8pt}
\paragraph{Use-case scenario.} Participants with a background in Computer Science, Digital Design and Development interact with two AI models: a Large Language Model and a Vision Language Model that supports video input. The goal is to assess whether these models can be utilized to train robots to understand observational videos depicting everyday interactions between people and objects. 

\subsection{Materials and methods}
We conducted a pilot workshop with experts in Design and Development to begin addressing the evident gap in user-centred VLM research. While the scale of this workshop limits the generalizability of our findings, it serves as an exploratory step toward identifying key challenges and opportunities for future user-involved studies. Our aim is not to offer definitive conclusions, but to begin collecting expert-grounded insights that foreground the importance of user involvement in this domain. 

We took previous work by Mehrotra et al. \cite{mehrotra2024integrity} as a reference and modelled trust decisions as users' choices to delegate tasks to an AI model. In this view, appropriate trust is achieved when the user's perceived trust in the system matches the actual system trustworthiness. That is, when users neither over-trust the system (over-rely on it, leading to misuse) nor under-trust it (under-rely on it, leading to disuse). This definition builds on users’ estimation of a system’s trustworthiness and is therefore more nuanced than directly measuring the system’s confidence in a given task. The two are related, as high system performance generally fosters higher trust. However, users may still perceive the system as more accurate than themselves yet choose to perform the task: for instance, using a manual gear shift in cars. Such cases are termed inconsistencies with respect to the system’s inherent trustworthiness \cite{mehrotra2024integrity}. Adopting the formal definitions of appropriate trust, over-trust, under-trust, and inconsistency in \cite{mehrotra2024integrity} entails simplifying assumptions: (i) tasks require exclusive decision-making, such that each is performed either by the AI or the user, without mixed cooperative scenarios; and (ii) appropriate trust is treated as a binary condition (true or false), requiring additional measures for trust calibration (e.g., a trust-meter scale). Nonetheless, this framework grounds high-level trust concepts in specific computer vision tasks, a necessary step for advancing the study of trust development in real-world experiments.

We structured workshop activities in two parts, each targeting one research question.
\vspace{-1em}
\paragraph{Preparatory activities.} Before beginning the two-part practical session, we conducted a 20-minute seminar to outline the broader objectives and context of the study. This session was important for ensuring participants shared a common understanding of: i) Vision Language Models and their distinction from Large Language Models, ii) the importance of assessing user trust in VLMs before deploying them in specific AI and Robotics scenarios, and iii) the diverse stakeholder groups to consider in future research. These range, depending on the downstream application, from industry and domain experts (e.g., for co-bots deployed in assembly lines) to the general public (e.g., in the case of household robots).

\begin{figure}[t]
    \centering
    \captionsetup{justification=centering}
    \includegraphics[width=\columnwidth, trim={20 20 30 10}, clip]{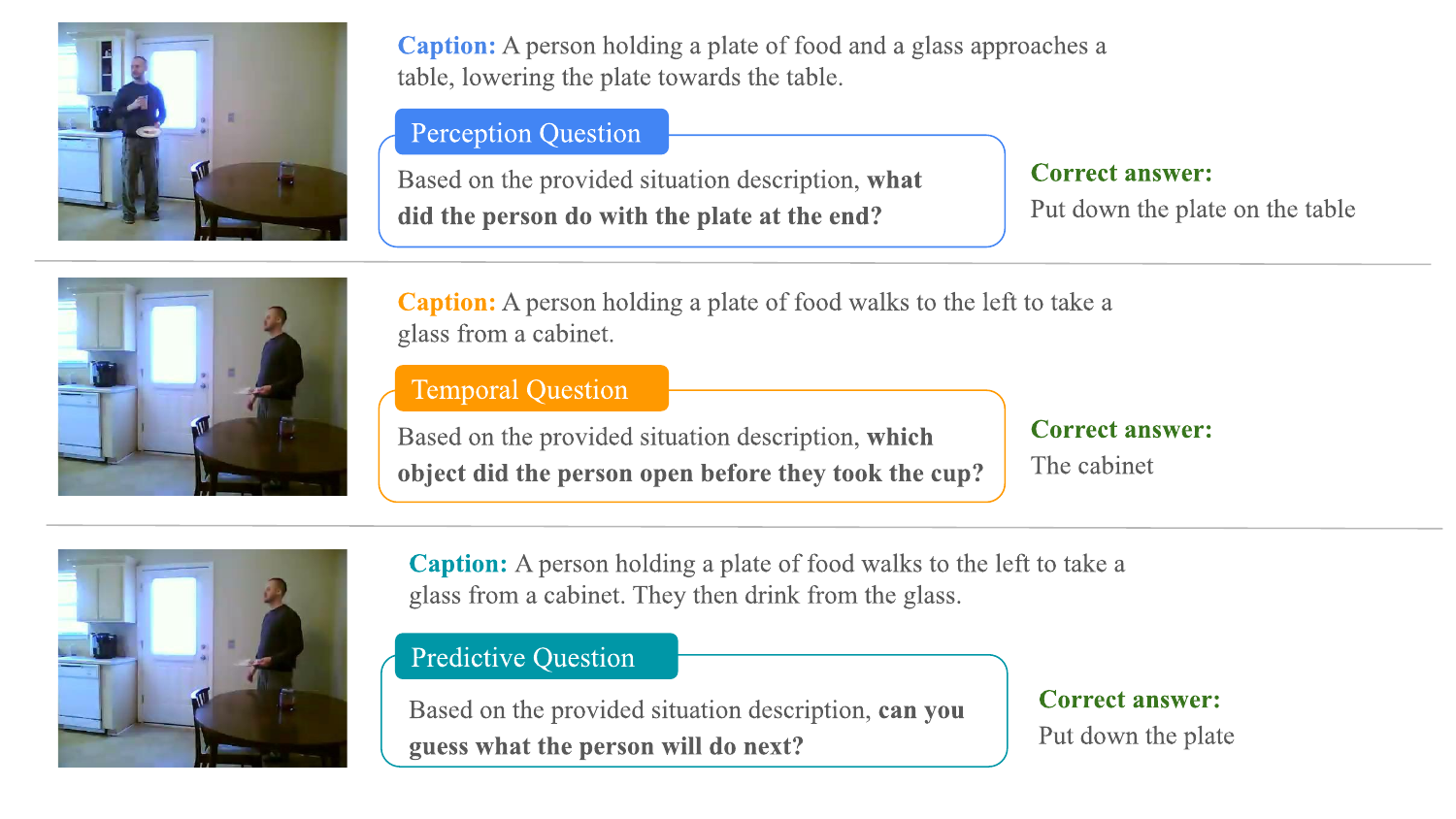} 
    \caption{Question examples within each task, by increasing difficulty.}
    \label{fig:star_examples}
\end{figure}
\vspace{-1em}

\begin{figure*}[t]
    \centering
     \captionsetup{justification=centering}
    \begin{subfigure}{\columnwidth} 
        \centering
    \includegraphics[width=0.9\textwidth,  trim={0 220 0 0}, clip]{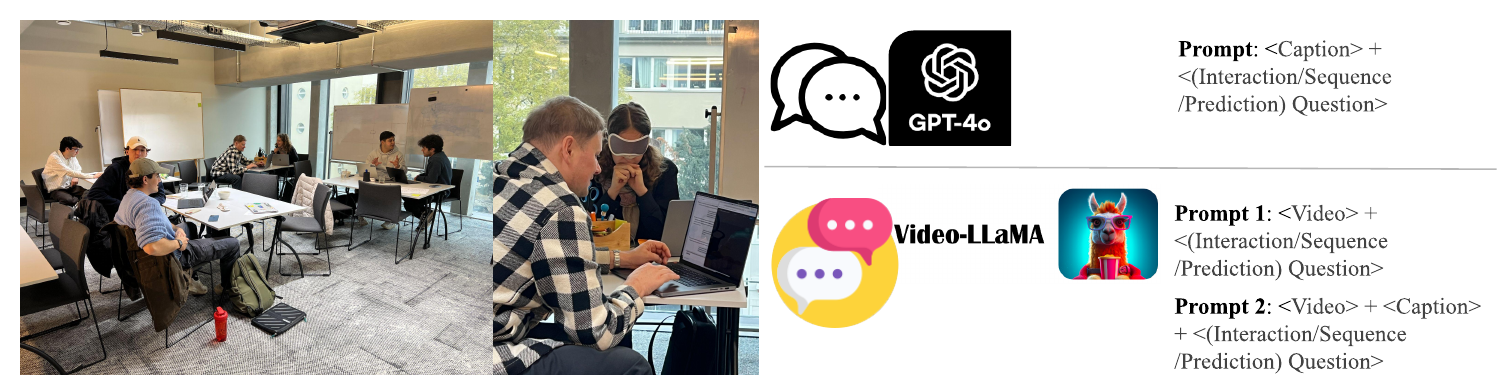} 
    \caption{ Players use different prompts to compare replies from the blindfolded player, the LLM, and the VLM.}
    \label{fig:collab_game}
    \end{subfigure}
    \hfill 
    \begin{subfigure}{\columnwidth} 
    \centering
    \includegraphics[width=\textwidth,  trim={0 0 0 180}, clip]{images/collaborative_game.pdf} 
    \caption{Question structure in the digital Miro board.}
    \label{fig:miro}
    \end{subfigure}
    \caption{Workshop Part 1 - The collaborative game.}
    \label{fig:focus_group}
\end{figure*}

\paragraph{Part 1: Collaborative game.} In the first phase, participants were paired to engage in a collaborative game. One participant was blindfolded, to be led to assess (and trust) only the words of the partner while deprived of other senses. Each  participant ($P1$) interacted with the blindfolded partner ($P2$) and an AI model to solve visual tasks derived from five situations in the STAR benchmark \cite{wu2021star}. Situations consisted of three video segments paired with three questions of increasing difficulty. This choice ensured to cover different levels of VLM cognitive ability \cite{wu2021star,liu_convbench_2024}: from the perception of actions involving people and objects (e.g., \textit{what did a person do...?}), to higher-level reasoning (e.g., \textit{what did the person do before/after X...?}), and forecasting future interactions (e.g., \textit{What will the person do next?}). Examples of question-answer pairs are provided in Figure \ref{fig:star_examples}.  
 One situation was used as a trial round, and four served as study tasks. 
 
 We structured the game as follows: $P1$ watched a video, read a textual description aloud to $P2$, and posed a question for $P2$ to answer. $P1$ then compared $P2$’s response to the correct answer and queried GPT-4o using the same textual description. This allowed the comparison between a blindfolded human and a language-only AI model. We supplemented STAR situations with textual descriptions written by us for this purpose. Both participants then interacted with Video-LLaMA 7B via \href{https://bit.ly/VLLaMa}{the Hugging Face Web demo}. Participants were instructed to always set the temperature parameter to the minimum value (0.1) to increase the model output consistency across trials. Two prompts were used: one included only the video attachment and question, while the other added the textual description (Figure \ref{fig:collab_game}). This setup enabled analysis of responses across multimodal and text-only inputs. We also encouraged players to modify the given prompts after the first interaction to test different prompting strategies. Participants recorded their observations on \href{https://miro.com/index/}{a digital Miro board}. The question structure is shown in Figure \ref{fig:miro}. 


\paragraph{Part 2: Mock-up evaluation.} In the second phase, participants reviewed graphical mock-ups of a proposed Web App designed to support a future extended study. The mock-ups are depicted in Figure \ref{fig:mockup}, where arrows represent the app navigation flow. 
Participants ranked features by importance (Figure \ref{fig:rank}), explained their rankings in free-text responses, and answered more detailed questions on individual features, through \href{https://www.mentimeter.com/}{Mentimeter}. An excerpt from the Mentimeter session is provided in Figure \ref{fig:menti}. Readers can refer to the ArXiV paper version \cite{chiatti2025mapping} for the complete question and rating structure used for this phase. Based on the survey findings, a key focus was evaluating whether a scene graph representation of the video was perceived as a useful addition to textual responses.


\subsection{Participants}
We held the workshop in December 2024 with 8 participants, organized into four teams for the collaborative game. To explore how trust develops in early interactions with a new model and data modality, specifically videos (\textbf{RQ1}), we recruited participants with varying familiarity with AI chatbots and LLMs but no significant experience with VLMs. Simultaneously, to gather feedback for designing the Web App (\textbf{RQ2}), we searched for participants with digital design and development skills. 
Participants were recruited at The CODE University of Applied Sciences in Berlin via a team collaboration platform among students and faculty members both well-versed in theoretical Software Engineering (SE) and Digital Design and Innovation (DDI) and actively engaged in real-world product design. 
Specifically, five students in SW and DDI, two Professors in DDI, and one researcher in participatory design took part, all proficient English speakers. Participants were fully informed about the study purpose and procedures through a detailed information sheet and formal consent form, which clarified that photos, group discussions, and activity outcomes would be recorded for academic purposes while maintaining confidentiality and anonymity of the provided data. The full consent form and information sheet is provided as appendix in the ArXiV paper version \cite{chiatti2025mapping}. Participation was voluntary, with the option to withdraw at any time; all participants completed the session.  

\subsection{Results}
\paragraph{Collaborative game.} After each task, players (1) compared the answers of the blindfolded partner with GPT-4o's responses (2) analysed Video-LLaMA's answers. We analysed player notes on Miro and re-watched the full workshop discussion recorded through a videoconferencing tool to summarise the results. 
\underline{For (1)}, the blindfolded player and GPT-4o achieved the same average ratio of correct answers across question types, i.e., 75\% accuracy. 
Interestingly, in some cases, the LLM could guess the correct answer while the blindfolded player could not, particularly for perception-level questions, where GPT-4o reached 100\% accuracy, surpassing the 91.67\% achieved by humans. 
Participants felt that the verbs chosen in the video description and questions (e.g., ``throwing'' vs. ``putting down'') were especially impactful on the players' understanding, leading some teams to test different verb synonyms in prompts. In temporal reasoning questions, GPT-4o accuracy dropped below the blindfolded players’ to 75\%, below the blindfolded players’ 83.33\%. 
As expected, predictive questions requiring causal reasoning and meta-learning proved hardest for both the LLM and participants, with both averaging 50\% accuracy. \\ Notably, in one case, both the human and GPT, despite providing the wrong answer, pointed out a discrepancy in object naming between the video caption and question that hindered comprehension. \\ 
\underline{For (2)}, querying Video-LLaMA with multimodal prompts led to the lowest accuracy for all question types: 29.63\% overall, 25\% on perception-level questions and 33.33\% on temporal reasoning questions. 
The model frequently hallucinated objects not present in the video, providing contradictory (``It said \texttt{[}the plate\texttt{]} was on the table at the end but also that the person was still holding it.'') 
and inconclusive answers (kept looping over the sentence ``the person is standing in a kitchen, and there is a white door in the background...'' without converging). 
Many noted that the VLM often avoided directly answering questions, regardless of the question asked, even when providing generally accurate descriptions of the video (``It didn't answer the question, it just described the video.''). 
Differently from GPT, the VLM relative performance was higher on temporal than on perception-level questions, likely due to being fine-tuned on temporally ordered frames. Predictive questions were again the hardest, with only 8.33\% accuracy. 
Video-LLaMA generally provided more conservative answers than GPT, especially on forecasting (``The model said it is not possible to determine what the person would eat'', the model replied ``We cannot determine whether they closed anything after holding the food because the action itself is not seen in the video''). 
Nonetheless, participants felt that GPT-4o provided ``more believable answers'', even when incorrect, compared to Video-LLaMA. Interestingly, one team experimented with using the VLM in text-only mode despite not being instructed to do so. While the model still struggled to answer correctly, adding textual context occasionally improved its guesses, reflecting the same language bias reported in related work \cite{gou2024eyes,huai_debiased_2024}. 

For future iterations of the study, we wrapped up the session asking participants \textit{What would you change about this game and why?} Participants expressed distrust in Video-LLaMA's capabilities, suggesting it be removed. They felt the game structure prioritised task completion over prompt refinement and successive interactions and recommended rephrasing prompts to allow for diverse verb forms. They also enjoyed the blindfold component, confirming that this strategy encouraged reliance on senses beyond vision. 

\begin{figure*}[t!]
    \begin{subfigure}{0.49\columnwidth} 
         \centering
    \includegraphics[scale=0.45, trim={10 60 80 10}, clip]{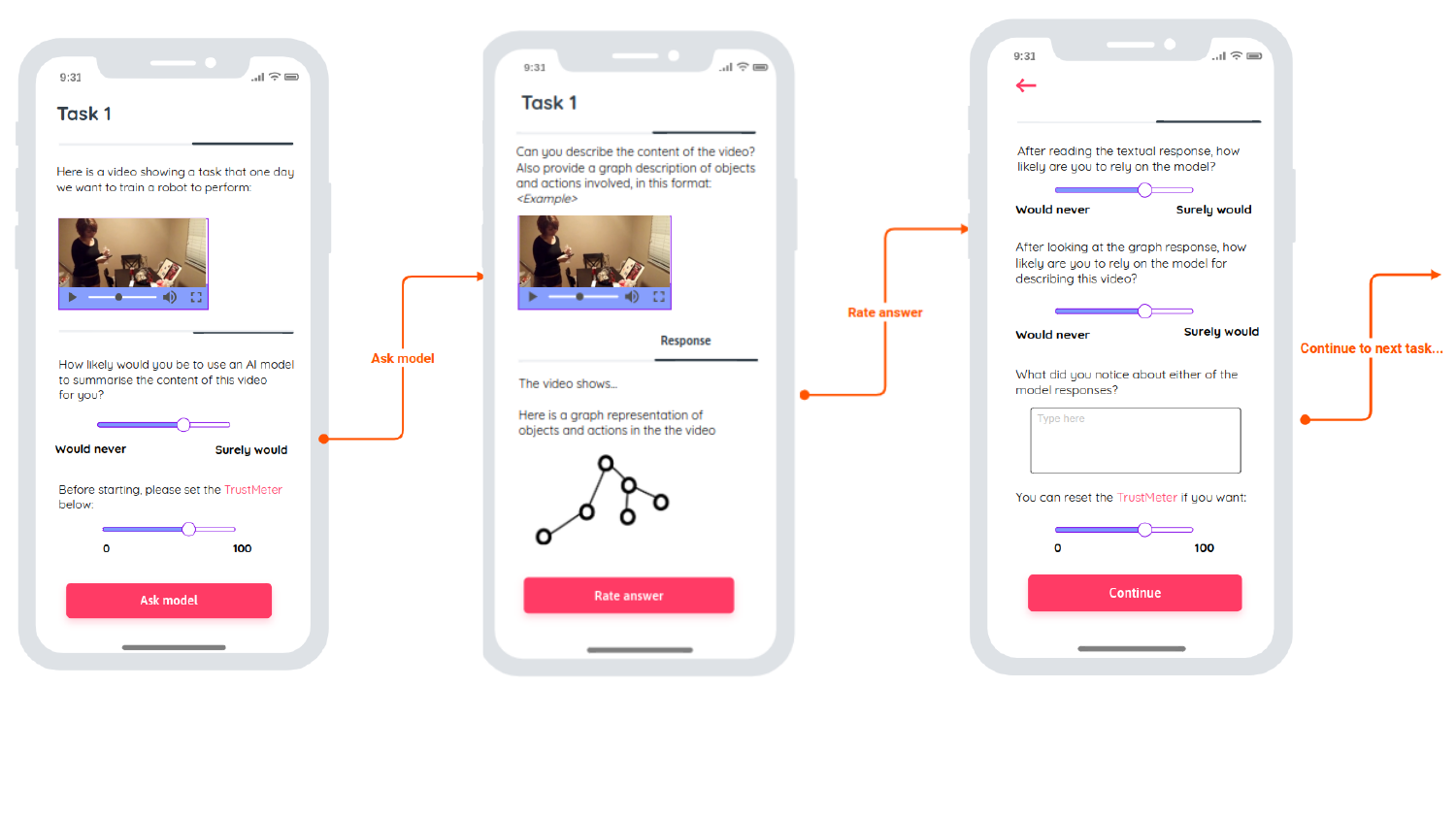} 
    \caption{App mock-ups under evaluation.  }
    \label{fig:mockup}
    \end{subfigure}
    \hfill
    \begin{subfigure}{0.5\columnwidth} 
         \centering
    \includegraphics[scale=0.37, trim={180 0 280 0}, clip]{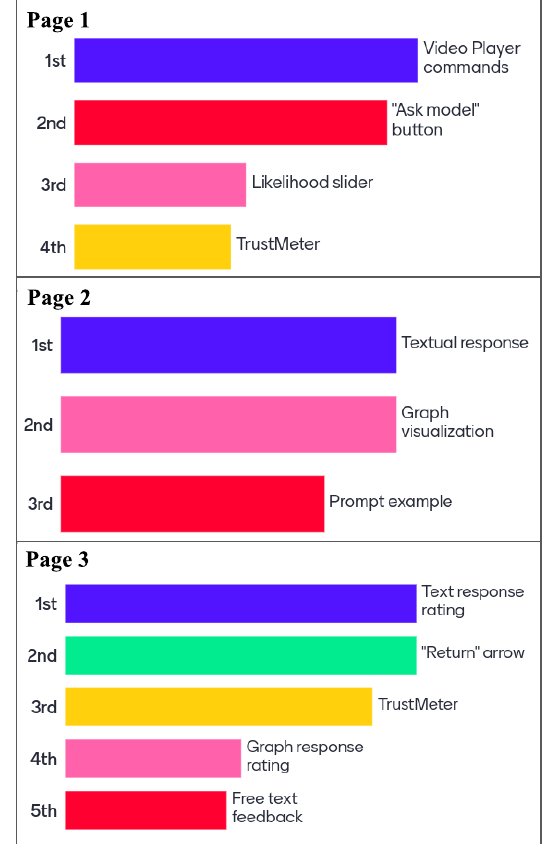} 
    \caption{Feature ranking exercise.}
    \label{fig:rank}
    \end{subfigure}
    \caption{Workshop Part 2 - The mock-up evaluation.}
    \label{fig:focus_group2}
\end{figure*}

\begin{figure*}
\centering
    \includegraphics[width=\textwidth,  trim={0 0 0 0}, clip]{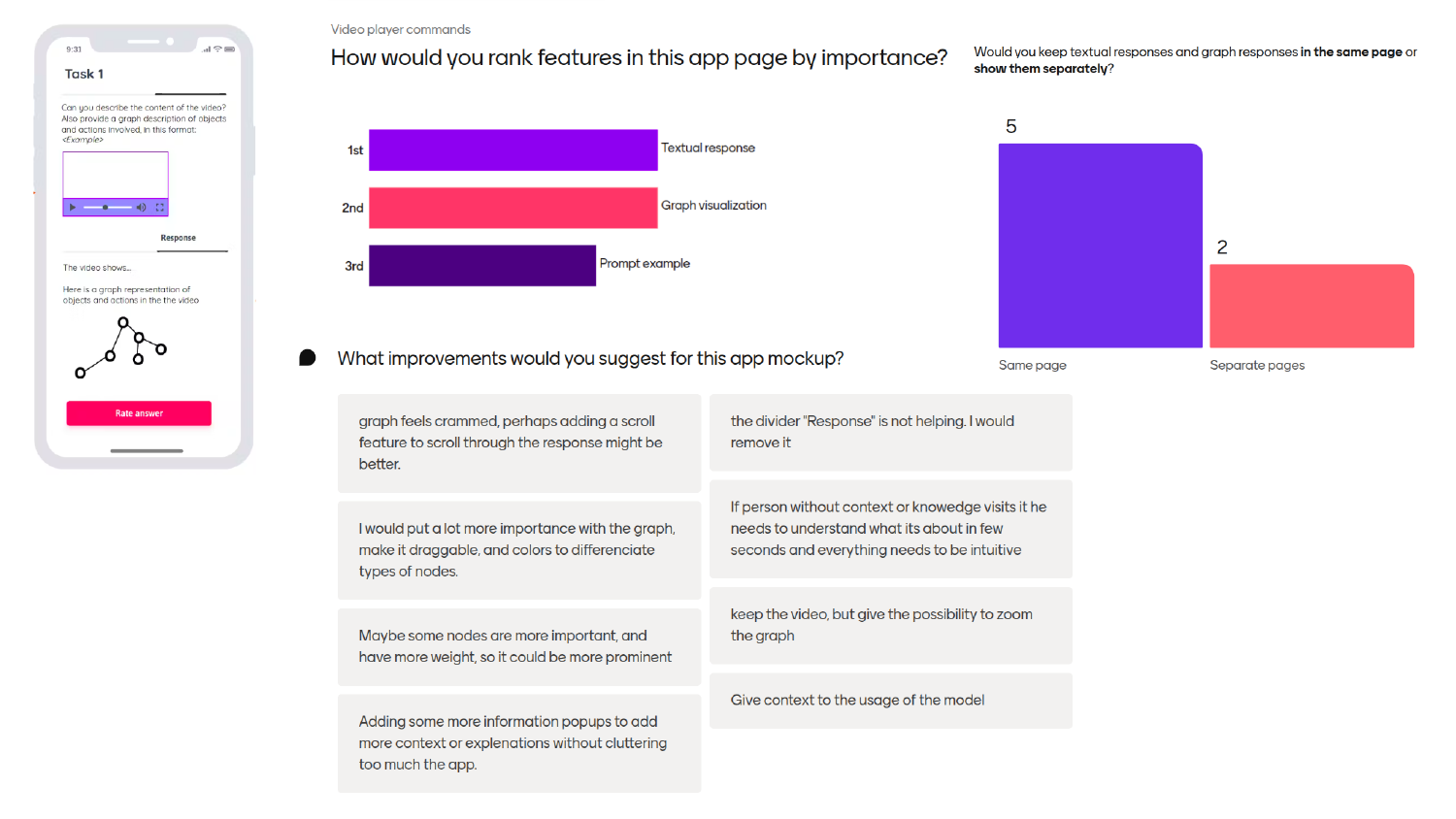} 
    \caption{Excerpt from the mock-up evaluation session via Mentimeter.} 
    \label{fig:menti}
\end{figure*}
\vspace{-1em}

\paragraph{Mock-up evaluation.} Participants' feature rankings are summarised in Figure \ref{fig:rank}. For \textbf{Page 1}, most users prioritised basic features over trust metric trackers like the TrustMeter, i.e., a feature we took from Mehrotra et al., \cite{mehrotra2024integrity} that allows users to set a score from 0 (distrust) to 100 (complete trust) to the model. Users preferred interacting with the model first before assigning a trust score and deemed these features less intuitive, especially for AI novices. 
They also suggested replacing continuous likelihood sliders with categorical ratings such as Likert-scale radio buttons. 
For \textbf{Page 2}, participants were presented with two app fields, one for evaluating the VLM's textual response and the other for evaluating the model graph-based response, i.e., the VLM output when asked to generate, in addition to the textual answer, a graph representing the observed scene. They rated textual and graph-based responses from the VLM as equally important, with 71\% preferring to display both on a single page (Figure \ref{fig:menti}).  
As shown in Figure \ref{fig:menti}, they suggested improvements like adding a scrolling sidebar, making the graph interactive (e.g., draggable, zoomable), and using colours to highlight node/edge types and importance. To enhance usability, they recommended explanatory pop-ups and removing unnecessary elements like the ``Response'' divider.
For \textbf{Page 3}, participants ranked the model text rating and the return button to revisit the responses highest (Figure \ref{fig:rank}). 
The TrustMeter was deemed more relevant on \textbf{Page 3} than on Page 1, while the graph rating and free-text feedback were seen as less important. Specifically, participants preferred annotating individual graph nodes/edges over rating the entire graph. While users felt trust scores were more likely to be completed than free-text input, they suggested requiring textual justifications in the case of extreme TrustMeter scores. They also suggested numbering rating questions for clarity. 
Overall, users found the app moderately intuitive and tasks easy to understand but noted that content balance on each page could be improved. While they were neutral about future use as they perceived no benefit beyond this research context, they felt the app could effectively support TAI studies and collaborative gamified evaluations. 
 
\section{Discussion and Future Work} 
This exploratory workshop, though limited in scale, provides initial insights into user trust dynamics when interacting with Vision-Language Models (VLMs). In the absence of established guidelines for evaluating user–VLM trust, this pilot discussion with experts helped identify preliminary design and methodological recommendations.
\vspace{-8pt}
\paragraph{Prioritising user agency over task completion.} Rather than focusing solely on task outcomes, future studies should enable users to engage in multi-turn interactions and iterative prompt refinement. This encourages deeper exploration of model behaviour and aligns with real-world usage patterns. Crucially, prioritising user agency in study design also contributes to enforcing human oversight in the interaction with the VLM. Moreover, while the workshop sample was small and homogeneous, discussions already pointed to the role of language choice (“throwing” vs. “putting down”) in shaping interpretations. In broader deployments, cultural, linguistic, and demographic differences may amplify bias and fairness concerns. Ethical evaluation must therefore consider diverse user groups to ensure VLM trust research does not replicate or reinforce systemic inequities.  
\vspace{-8pt}
\paragraph{Contextualizing trust metrics.} Tools such as the TrustMeter should be introduced alongside clear explanations and framed within concrete research objectives. Without this context, these features risk being perceived as arbitrary or uninformative, negatively impacting user engagement and inclusion.

\vspace{-8pt}
\paragraph{Exploring graph responses.} In addition to textual and visual outputs, the integration of structured representations such as scene graphs can enable collecting more granular feedback and support user comprehension. Our findings suggest that participants valued the hybrid nature of graphs, especially when paired with interactive and visual enhancements (e.g., zoom, drag, colour-coded nodes). As such, graphs could provide an interpretable interface to help assign accountability in cases of error, as well as enhance responsible human oversight. 

\vspace{-8pt}
\paragraph{Tracking and calibrating trust.} As demonstrated by user reactions to Video-LLaMA hallucinations and non-responsiveness, trust can degrade sharply after a few poor experiences. Our results show that participants sometimes perceived VLM outputs as “believable” even when inaccurate. This raises concerns about over-trust, where users may defer to system outputs despite errors. We also found that some participants distrusted Video-LLaMA after a few poor responses. While healthy skepticism is important, systematic under-trust could lead to disuse of potentially beneficial tools, particularly by groups with less exposure to AI technologies. Thus, from an ethical perspective, designing equitable trust calibration mechanisms is a priority. The continuous tracking of trust metrics throughout the evaluation session is also essential, as opposed to a single post-hoc rating. However, trust scores can be hard to assess a priori and ice-breaking tasks could help users calibrate their early-stage assessments.

\vspace{-8pt}
\paragraph{Gamification and engagement strategies.} The blindfolded game element was well received and viewed as an effective means to shift focus to verbal and situational cues. Building on this insight, future tools might incorporate gamified tasks or collaborative challenges to enhance user engagement and capture richer behavioural data when evaluating trust. 

Building upon these observations and expanding on the limitations of our pilot workshop, we outline several directions for improvement in future research.
\vspace{-8pt}
\paragraph{Longitudinal and in-situ evaluation.} Trust evolves dynamically over time and across contexts. Future studies should prioritize longitudinal designs to observe how trust is built, maintained, or eroded over extended use. This could involve the in-field deployment of VLM-enabled tools and repeated evaluation sessions.
\vspace{-8pt}
\paragraph{Participant diversity and inclusion.} It is critical to recruit a demographically diverse user base, spanning different levels of AI literacy, cultural backgrounds, and application domains. In particular, the inclusion of underrepresented and vulnerable groups is vital to understand how implicit biases in model outputs affect trust.

\vspace{-8pt}
\paragraph{Context-specific use cases.} While our workshop focused on general-purpose videos requiring situational reasoning, future studies may compare trust dynamics across lower-risk and higher-risk scenarios, the latter including, for instance, healthcare, legal decision support, and education.
\vspace{-8pt}
\paragraph{Model benchmarking and comparison.} Incorporating multiple VLMs, including domain-tuned variants, can help disentangle model-specific behaviours and flaws from more general user trust patterns. This comparative approach is especially important given the variability in VLM performance across modalities and tasks.
\vspace{-8pt}
\paragraph{Trust repair and recovery mechanisms.} Little is known about how trust in VLMs can be rebuilt after failure. Studies should investigate whether feedback loops, corrective interactions, or transparency about model limitations can facilitate trust recovery, especially in mixed-initiative systems. 

 Although research on trust dynamics in human–VLM cooperation is still in its early stages, the outlined trajectories and requirements provide a useful starting point for researchers, practitioners, and policymakers to advance the study of user–VLM trust. They lay the groundwork for more extensive and targeted user studies aimed at deriving design principles for specific applications, contributing to efforts to align regulatory frameworks with real-world deployment.   

\begin{acknowledgments}
  This work has been supported by Politecnico di Milano through the 2024 MSCA Seal of Excellence fellowship (project ReFiNe) and by the FAIR (Future Artificial Intelligence Research) project, funded by the NextGenerationEU program within the PNRR-PE-AI scheme (M4C2, investment 1.3, line on Artificial Intelligence).  
\end{acknowledgments}

\section*{Declaration on Generative AI}
 During the preparation of this work, the author(s) used GPT-4-turbo for grammar and spell checks. 

\bibliography{ecai25}


\end{document}